\documentclass[twocolumn,secnumarabic,amssymb, amsmath, nofootinbib,tightenlines,
nobibnotes, aps, prl,epsfig]{revtex4}
\usepackage{graphicx}
\usepackage{dcolumn}
\usepackage{bm}
\begin{document}
\preprint{APS/123-QED}
\title{The universal function of the diffractive process in color dipole picture }
\author{Z. Jalilian }
\email{zeinab.jalilian.kr@gmail.com}
\affiliation{Department of Physics, Razi University, Kermanshah, 67149, Iran}
\author{G. R. Boroun }
\email{boroun@razi.ac.ir, grboroun@gmail.com}
\affiliation{Department of Physics, Razi University, Kermanshah, 67149, Iran}
\date {\today}

\begin{abstract}
In this presentation, we obtain the corresponding universal function to the diffractive process and show the cross section exhibits the geometrical scaling.
 It is observed the diffractive theory according to the color dipole approach at small-$x$ is a convenient framework that reveals the color transparency
 and the saturation phenomena. Also we calculate the contribution of heavy quark productions in the diffractive cross section for high energy that is
 determined by the small size dipole configuration. The ratio of the diffractive cross section to the total cross section in the electron-proton collision
 is the other important quantity that is computed in this work.\\

\end{abstract}
 \pacs{***}
\maketitle
\section {\textbf{1. Introduction}}

The color dipole formalism in QCD prediction for high energy deep inelastic scattering at small-x has promoted a lot of
phenomenological activity successfully, in recent years. The saturation effect in $x<0.01$ dominates by the gluon dynamics
that describes the details of the electron-proton collision data collected at HERA well [1-4].\\
Understand the diffractive deep inelastic scattering is a great theoretical feature because 10
to 15 percent of all events observed at HERA are diffractive [5-7]. Many of the experimental data can be explained by perturbative QCD,
 but an extrapolation to diffractive reactions must carefully be performed because the most of them are sensitive to details of non-perturbative dynamics.
The study on the diffraction process cames from pioneering work of Glauber [8] that developed by Good and Walker as a quantum mechanical effect [9].\\
Indeed, at small-$x$ a diffractive process in DIS, in the electron-proton collision, occurs in the form of $ eP \longrightarrow eXP$.
The dynamics behind this event is simply justified if we review it in the rest frame of the proton. In this case the target proton remains intact,
a photon with virtuality $Q^{2}$ develops a partonic fluctuation and is forced to a strong interaction with the proton as
 $\gamma^{\star}p \longrightarrow Xp$, so a large rapidity gap (LRG) appears between the scattered proton and particle flow
  formed from the virtual photon in the final state.\\
 From the perspective of the color dipole approach we can express: when $x \longrightarrow 0$, in the rest frame of the target,
  the virtual photon splits up to a quark-antiquark pair before the scattering tagged with Fock eigenstate $\vert q\overline{q}\rangle $.
  This eigenstate is expressed by the quantum mechanical wave function with probability
\begin{eqnarray}
 \nonumber \vert \Psi_{T}(z,r) \vert ^{2}&=& _{T}\langle q\overline{q}\vert q\overline{q}\rangle _{T}
 \\
 \nonumber &=&\dfrac{N_{c}\alpha_{em}}{2\pi^{2}}\sum_{q} e_{q}^{2} [(z^{2}+(1-z)^{2})\varepsilon^{2}K_{1}^{2}(\varepsilon r)
 \\ &+&m_{q}^{2}K_{0}^{2}(\varepsilon r)],
\end{eqnarray}
for the transversely and\\
\begin{eqnarray}
 \nonumber \vert \Psi_{L}(z,r) \vert ^{2}&=&_{L}\langle q\overline{q}\vert q\overline{q}\rangle _{L}
  \\
 \nonumber &=&\frac{N_{c}\alpha_{em}}{2\pi^{2}}\sum_{q} e_{q}^{2}
  4 Q^{2}z^{2}(1-z)^{2} K_{0}^{2}(\varepsilon r),
   \\
\end{eqnarray}
for the longitudinally polarized photon.\\
In above-mentioned equations the fraction of the momentum carrying by quark, $z$, and the relative transverse separation of the $q\overline{q}$ pair, $r$,
are appropriate freedom degrees. The contribution of each quark flavor is proportional to its electromagnetic charge and inverse mass,
 $N_{c}=3$, $ \varepsilon ^{2}=z(1-z)Q^{2}+(m_{q})^{2} $ and  the center of mass energy squared of $\gamma^{\star}p$ is $W^{2}$ that
  $x=\dfrac{Q^{2}+4m_{q}^{2}}{W^{2}}$. Also for $ \varepsilon r < 1 $, $K_{1,0}(\varepsilon r)$ is estimated by MacDonald Bessel functions [10].\\
  These equations include Gribov inelastic shadowing corrections to all of multiple interactions what is hardly possible within hadronic presentation [11].
For a dipole-proton interaction we use the dedicated  cross section formulated by Bartels, Golec-Bierant and Kowalski as a suitable
 definition which involves the gluon distribution function [12]
\begin{eqnarray}
\nonumber \sigma_{q\overline{q}P}(x,r^{2} ) = \sigma_{0} \Big \{
1-\exp(-\dfrac{
\pi^{2}r^{2}\alpha_{s}(\mu^{2})xg(x,\mu^{2})}{3\sigma_{0}}) \Big
\},
\\
\end{eqnarray}
with $\mu^{2}=\mu_{0}^{2}+\dfrac{C}{r^{2}}$ that parameters $C$ and $\mu_{0}^{2}$ from fit to DIS data are be determined.
 The dipole-hadron cross section $\sigma_{q\overline{q}P}$ contains information about  the strong interaction physics and the target.
In fact, we note the polarized photon in addition to the transverse and the longitudinal quark-antiquark pairs can be spilt up to a
transverse $q\overline{q}g$ dipole dominated in final state due to the gluon production. We know the importance of this contribution that
 has been studied in references [13,14] for nucleon and nucleus, of course, in a different way than our method.
  Since the $q\overline{q}g$ dipole is created by assuming a strong ordering in the transverse space $r$ the
   fraction of the momentum carrying by the gluon comparing with the corresponding value for the quark and antiquark
    is much smaller and we can ignore it. In the other word the most of the energy is carried by the hadron and the
    virtual photon has just enough energy to dissociate into a $q\overline{q}$ pair before the scattering. Thus the diffractive
     deep inelastic scattering cross section is formulated as
\begin{eqnarray}
 \int_{-\infty}^{0} dt e^{B_{D}t}\dfrac{d\sigma^{D}_{T,L}}{dt}\vert_{t=0}&=&\dfrac{1}{B_{D}}\dfrac{d\sigma^{D}_{T,L}}{dt}\vert_{t=0},
\end{eqnarray}
by considering a factorization dependence on $t$ with the
diffractive slope $B_{D}$ [15]. Where
\begin{eqnarray}
\nonumber \dfrac{d\sigma^{D}_{T,L}}{dt}\vert_{t=0}&=&\dfrac{1}{16\pi} \Big (\langle \sigma^{2}_{q\overline{q}P}(x,r^{2}) \rangle_{T,L} - \langle \sigma_{q\overline{q}P}(x,r^{2}) \rangle^{2}_{T,L} \Big ).
\\
\end{eqnarray}
The definition of the expectation value is
\begin{eqnarray}
\nonumber \langle \sigma _{q\overline{q}P}(x,r^{2}) \rangle_{T,L}&=&_{L,T}\langle q\overline{q} \vert \sigma_{q\overline{q}P} (x,r^{2}) \vert q\overline{q} \rangle_{T,L},
\\
\nonumber &=&\sigma_{T,L}^{\gamma^{\star}P}(x,Q^{2}),
\\
\nonumber &=& \int_{0}^{1}dz\int d^{2}r  \vert \Psi_{T,L}(z,r)\vert ^{2} \sigma_{q\overline{q}P}(x,r^{2}).
\\
\end{eqnarray}
Since $\langle \sigma_{q\overline{q}P}(x,r^{2}) \rangle_{T,L}=O(\alpha_{em})$ we can ignore the second term of Eq. (5)
in comparison to the first one and hence we will obtain\\
\begin{eqnarray}
\nonumber \dfrac{d\sigma^{D}_{T,L}}{dt}\vert_{t=0}&=&\dfrac{1}{16\pi} \Big (\langle \sigma^{2}_{q\overline{q}P}(x,r^{2}) \rangle_{T,L}\Big ),
\\
\nonumber &=& \dfrac{1}{16\pi}\int_{0}^{1}dz\int d^{2}r  \vert \Psi_{T,L}(z,r)\vert ^{2} \sigma^{2}_{q\overline{q}P}(x,r^{2}).
\\
\end{eqnarray}
That's mean the diffractive cross section is a quantum mechanics summation over the effective dipole cross section square,
 $\sigma^{2}_{q\overline{q}P}(x, r^{2})$, for different Fock states [16].\\
The content of this paper is the following. In section 2 we
calculate the diffractive cross section and the universal function
and investigate existence of the geometrical scaling. Since the presence of the heavy pairs in the high energy is important we
obtain their contribution in section 3. Also in this section the ratio of the diffractive cross section to the
 total cross section is determined by the ratio of the corresponding universal functions. Finally results in section 4 are summarized.\\

\section {\textbf{2. the diffractive universal function}}

Before computing, we introduce the $x$-dependence saturation radius, $R_{0}(x)$, related to the saturation scale, $Q_{s}$,
is a energy-dependent scale and is a critical element in determining the saturation point
\begin{eqnarray}
R_{0}^{2}(x)=\dfrac{1}{Q_{0}^{2}}(\dfrac{x}{x_{0}})^{\lambda}=\dfrac{1}{Q_{s}^{2}}.
\end{eqnarray}
Positive and constant variables $Q_{0}$, $x_{0}$ and $\lambda$ have been obtained by fitting done with H1 and ZEUS data
 by Golec-Bierant and W$\mathrm{\ddot{u}}$thoff [17].\\
According to Eq. (3) the selection of the gluon distribution is important. For small-$r$, $r<R_{0}(x)$, the gluon distribution is modelled as
 \begin{eqnarray}
xg(x,\mu^{2})=\dfrac{3\sigma_{0}}{4\pi^{2}\alpha_{s}R_{0}^{2}(x)},
\end{eqnarray}
where $\mu^{2}$ behaves as $\dfrac{C}{r^{2}}$. Then for small-$r$
\begin{equation}
\sigma_{q\overline{q}P}(x,r^{2} )=\sigma_{0}\dfrac{r^{2}}{R_{0}^{2}(x)}.
\end{equation}
For larg-r the scale $\mu^{2}$ is closed to $\mu_{0}^{2}$ and according to the original saturation model the saturation value of
the dipole cross section is $\sigma_{q\overline{q}P}(x,r^{2} )\approx \sigma_{0}$ [18, 19].
We can summarize the contents as
\begin{equation}
\sigma_{q\overline{q}P}(x,r^{2} )=\Bigg\{
\begin{array}{cc}
   \sigma_{0},&r>R_{0}\\
     \sigma_{0}\dfrac{r^{2}}{R_{0}^{2}},&r<R_{0}
\end{array}
\end{equation}
that we have obtained the saturation cross section, $\sigma_{0}$, in agreement with data reported by H1 and ZEUS, for more details see Ref [20].\\
Now we are ready to determine the diffractive cross section. Since the $q\overline{q}$ pairs with
the size $r^{2}  \sim \dfrac{1}{\varepsilon^{2}} \cong \dfrac{1}{Q^{2} z(1-z)}$ make the dominant contribution
we need to solve the integral of Eq. (7) for $ \varepsilon r < 1 $ with
$0 \leq z \leq 1$ .\\
There are two limit states which are interesting to investigate:
one of them is the symmetric pairs with $ r \leq 1/Q $ in this
case the quark and antiquark carry the equal contribution of the
photon transverse momentum. The size of this type of the color
dipole
 is small in comparison to the saturation radius, $r<R_{0}$, by substituting relations (1), (2) and (11) in Eq.(7) we can obtain
 the summation over the transverse and longitudinal contributions\\
\begin{eqnarray}
\nonumber \dfrac{d\sigma^{D}_{tot}}{dt}\vert_{t=0}&=&\dfrac{d\sigma^{D}_{T}}{dt}\vert_{t=0}+\dfrac{d\sigma^{D}_{L}}{dt}\vert_{t=0} ,
\\
\nonumber &=&\dfrac{3\alpha_{em}}{32\pi^{3}}\sigma^{2}_{0}\sum_{q} e_{q}^{2}\Big \{ \int_{0}^{1}dz(z^{2}+(1-z)^{2})
\\
\nonumber &\times & \int_{0}^{\dfrac{1}{Q^{2}}}d^{2}r\varepsilon^{2}(\dfrac{1}{\varepsilon^{2}r^{2}})(\dfrac{r^{2}}{R_{0}^{2}(x)})^{2}
\\
\nonumber &+&\int_{0}^{1}dz\int_{0}^{\dfrac{1}{Q^{2}}}d^{2}r m_{q}^{2}(\dfrac{r^{2}}{R_{0}^{2}(x)})^{2}
\\
\nonumber &+&\int_{0}^{1}dz 4Q^{2}z^{2}(1-z)^{2}
\\
\nonumber &\times &  \int_{0}^{\dfrac{1}{Q^{2}}}d^{2}r(\dfrac{r^{2}}{R_{0}^{2}(x)})^{2} \Big \},
\\
\nonumber &=&\dfrac{3\alpha_{em}}{32\pi^{2}}\sigma^{2}_{0}\dfrac{1}{3Q^{4}R_{0}^{4}(x)}\sum_{q} e_{q}^{2}\big \{\dfrac{17}{15}+\dfrac{m_{q}^{2}}{Q^{2}} \big \}.
\\
\end{eqnarray}
We see the diffractive cross section is as small as $\dfrac{1}{Q^{4}}$ therefore the main contribution comes from rare
 fluctuations of photon that corresponds to the color transparency configuration which happens rarely. \\
The idea of the geometrical scaling has been based on writing the total cross section as a function of the dimensionless variable $\tau$ as [15]
\begin{eqnarray}
\sigma_{T,L}^{\gamma^{\star}P}(x,Q^{2})=\sigma_{0}f(\tau).
\end{eqnarray}
In this work, we select the scaling variable as $\tau = R_{0}^{2}(x)Q^{2}$ and generalize this idea to the diffractive cross section and write in a similar way
\begin{equation}
 \dfrac{d\sigma^{D}_{tot}}{dt}\vert_{t=0}=\sigma^{2}_{0}g(\tau),
\end{equation}
Then Eq. (12) is rewritten as
\begin{eqnarray}
\nonumber \dfrac{d\sigma^{D}_{tot}}{dt}\vert_{t=0}&=&\dfrac{3\alpha_{em}}{32\pi^{2}}\sigma^{2}_{0}\dfrac{1}{3\tau^{2}}\sum_{q} e_{q}^{2}\big \{ \dfrac{17}{15}+\dfrac{m_{q}^{2}}{Q^{2}}\big \}.
\\
\end{eqnarray}
that
\begin{equation}
g(\tau)=\dfrac{3\alpha_{em}}{32\pi^{2}}\dfrac{1}{\tau^{2}}\sum_{q} e_{q}^{2}\big \{ \dfrac{17}{45}+\dfrac{m_{q}^{2}}{3Q^{2}}\big \}.
\end{equation}
To continue we obtain the total diffractive cross section where the  dipole size is larger than the saturation radius
\begin{eqnarray}
\nonumber \dfrac{d\sigma^{D}_{tot}}{dt}\vert_{t=0}&=&\dfrac{d\sigma^{D}_{T}}{dt}\vert_{t=0}+\dfrac{d\sigma^{D}_{L}}{dt}\vert_{t=0},
\\
\nonumber &=&\dfrac{3\alpha_{em}}{32\pi^{3}}\sigma^{2}_{0}\sum_{q} e_{q}^{2}\Big \{ \int_{0}^{1}dz(z^{2}+(1-z)^{2})
\\
\nonumber &\times & \int_{0}^{R_{0}^{2}}d^{2}r\varepsilon^{2}(\dfrac{1}{\varepsilon^{2}r^{2}})(\dfrac{r^{2}}{R_{0}^{2}(x)})^{2}
\\
\nonumber &+&\int_{0}^{1}dz\int_{0}^{R_{0}^{2}}d^{2}r m_{q}^{2}(\dfrac{r^{2}}{R_{0}^{2}(x)})^{2}
\\
\nonumber &+&\int_{0}^{1}dz(z^{2}+(1-z)^{2}) \int_{R_{0}^{2}}^{\dfrac{1}{Q^{2}}}d^{2}r\varepsilon^{2}(\dfrac{1}{\varepsilon^{2}r^{2}})
\\
\nonumber &+&\int_{0}^{1}dz\int_{R_{0}^{2}}^{\dfrac{1}{Q^{2}}}d^{2}r m_{q}^{2}
\\
\nonumber &+&\int_{0}^{1}dz 4Q^{2}z^{2}(1-z)^{2} \int_{0}^{R_{0}^{2}}d^{2}r(\dfrac{r^{2}}{R_{0}^{2}(x)})^{2}
\\
\nonumber &+&\int_{0}^{1}dz 4Q^{2}z^{2}(1-z)^{2} \int_{R_{0}^{2}}^{\dfrac{1}{Q^{2}}}d^{2}r \Big \},
\\
\nonumber &=&\dfrac{3\alpha_{em}}{32\pi^{2}}\sigma^{2}_{0}\sum_{q} e_{q}^{2}\big \{ \dfrac{1}{3}(\dfrac{7}{5}-log(R_{0}^{2}Q^{2})^{2}
\\
\nonumber &-&\dfrac{4R_{0}^{2}Q^{2}}{15})+\dfrac{m_{q}^{2}}{Q^{2}}(1- \dfrac{2R_{0}^{2}Q^{2}}{3}) \big \}.
\\
\end{eqnarray}
$g(\tau)$ in this case by ignoring the logarithmic sentence is expressed by
\begin{eqnarray}
\nonumber g(\tau)&=&\dfrac{3\alpha_{em}}{32\pi^{2}}\sum_{q} e_{q}^{2}\big \{ \dfrac{7}{15}-\dfrac{4\tau}{45}+\dfrac{m_{q}^{2}}{Q^{2}}(1- \dfrac{2\tau}{3}) \big \}.
\\
\end{eqnarray}
This function when $\tau \longrightarrow 0$ becomes
\begin{eqnarray}
\nonumber g(\tau)&=&\dfrac{3\alpha_{em}}{32\pi^{2}}\sum_{q} e_{q}^{2}\big \{ \dfrac{7}{15}+\dfrac{m_{q}^{2}}{Q^{2}} \big \}.
\\
\end{eqnarray}
therefore
\begin{eqnarray}
g(\tau) \cong O(\alpha_{em}).
\end{eqnarray}
We have plotted the ratio $\dfrac{g(\tau)}{\alpha_{em}}$ for Eqs.
(16) and (18) in terms of $\tau$ variable in Fig. 1 in different
$x$ values for light flavors with $m_{q} = 140~\mathrm{MeV}$.
According to these diagrams $\tau = 1$ divides the plane to the
saturation and scaling areas and all of them are independent of
$x$. Also we see the slope each diagram in the scaling region is
steeper in comparison to the corresponding universal function to
the total cross section, for more information see Ref. [20]. We
can briefly express if $r$ changes from $r > R_{0}$ to $r < R_{0}$
the unitarity effect in $ 0 < \tau < 1 $ region links to a weak
interaction in $\tau \geq 1$, on the other hand the universal
function in Fig. 1 behaves as the following
\begin{eqnarray}
\dfrac{g(\tau)}{\alpha_{em}} \sim 1 \longrightarrow \dfrac{g(\tau)}{\alpha_{em}} \sim \dfrac{1}{\tau^{2}}.
\end{eqnarray}
The other limit state occurs when one of the components of the pair carries a large part of the transverse momentum. The color dipole
created in this case is called the asymmetric pair. We note the condition $\varepsilon r<1$  in Eq. (7) is fulfilled only
 if $ z<\dfrac{1}{r^{2}Q^{2}}$ also there must be a cut-off such as $\mu^{2} \simeq 4m_{q}^{2}$ on the energy. Where
  the asymmetric dipole size is smaller than the saturation radius we will have
\begin{eqnarray}
\nonumber \dfrac{d\sigma^{D}_{tot}}{dt}\vert_{t=0}&=&\dfrac{d\sigma^{D}_{T}}{dt}\vert_{t=0}+\dfrac{d\sigma^{D}_{L}}{dt}\vert_{t=0},
\\
\nonumber &=&\dfrac{3\alpha_{em}}{32\pi^{3}}\sigma^{2}_{0}\sum_{q} e_{q}^{2}\Big \{
 \int_{1/\mu^{2}}^{1/Q^{2}}d^{2}r\varepsilon^{2}(\dfrac{1}{\varepsilon^{2}r^{2}})
\\
\nonumber &\times & \int_{0}^{\dfrac{1}{r^{2}Q^{2}}}dz(z^{2}+(1-z)^{2})(\dfrac{r^{2}}{R_{0}^{2}(x)})^{2}
\\
\nonumber &+&  \int_{1/\mu^{2}}^{1/Q^{2}}d^{2}r\int_{0}^{\dfrac{1}{r^{2}Q^{2}}}dz 4Q^{2}z^{2}(1-z)^{2}(\dfrac{r^{2}}{R_{0}^{2}(x)})^{2} \Big \},
\\
\nonumber &+&\int_{1/\mu^{2}}^{1/Q^{2}}d^{2}r m_{q}^{2}\int_{0}^{\dfrac{1}{r^{2}Q^{2}}}dz(\dfrac{r^{2}}{R_{0}^{2}(x)})^{2}
\\
\nonumber &=&\dfrac{3\alpha_{em}}{32\pi^{2}}\sigma^{2}_{0}\dfrac{\sum_{q} e_{q}^{2}}{Q^{4}R_{0}^{4}(x)}\big \{ \dfrac{29}{15}+\dfrac{1}{3}log(\dfrac{\mu^{2}}{Q^{2}})
\\
\nonumber &-& \dfrac{4\mu^{2}}{3Q^{2}}+\dfrac{2\mu^{4}}{5Q^{4}}+\dfrac{m_{q}^{2}}{2Q^{2}}(1-\dfrac{Q^{4}}{\mu^{4}}) \big \}.
\\
\end{eqnarray}
According to the high power of the virtuality, this function falls much faster than the corresponding case in small pairs
 and the interaction is almost unexpected. The universal function in this case is given by
\begin{eqnarray}
\nonumber g(\tau)&=&\dfrac{3\alpha_{em}}{32\pi^{2}}\dfrac{1}{\tau^{2}}\sum_{q} e_{q}^{2}\big \{ \dfrac{29}{15}+\dfrac{1}{3}log(\dfrac{\mu^{2}}{Q^{2}})
\\
\nonumber &-& \dfrac{4\mu^{2}}{3Q^{2}}+\dfrac{2\mu^{4}}{5Q^{4}}+\dfrac{m_{q}^{2}}{2Q^{2}}(1-\dfrac{Q^{4}}{\mu^{4}}) \big \}.
\\
\end{eqnarray}
Finally we investigate the asymmetric pair where its size is large in comparison to the saturation radius
\begin{eqnarray}
\nonumber \dfrac{d\sigma^{D}_{tot}}{dt}\vert_{t=0}&=&\dfrac{d\sigma^{D}_{T}}{dt}\vert_{t=0}+\dfrac{d\sigma^{D}_{L}}{dt}\vert_{t=0},
\\
\nonumber &=&\dfrac{3\alpha_{em}}{32\pi^{3}}\sigma^{2}_{0}\sum_{q} e_{q}^{2}\Big \{
 \int_{1/ \mu^{2}}^{R_{0}^{2}}d^{2}r\varepsilon^{2}(\dfrac{1}{\varepsilon^{2}r^{2}})
\\
\nonumber &\times &\int_{0}^{\dfrac{1}{r^{2}Q^{2}}}dz(z^{2}+(1-z)^{2})(\dfrac{r^{2}}{R_{0}^{2}(x)})^{2}
\\
\nonumber &+&\int_{1/ \mu^{2}}^{R_{0}^{2}}d^{2}r m_{q}^{2}\int_{0}^{\dfrac{1}{r^{2}Q^{2}}}dz(\dfrac{r^{2}}{R_{0}^{2}(x)})^{2}
\\
\nonumber &+& \int_{R_{0}^{2}}^{1/Q^{2}}d^{2}r\varepsilon^{2}(\dfrac{1}{\varepsilon^{2}r^{2}})\int_{0}^{\dfrac{1}{r^{2}Q^{2}}}dz(z^{2}+(1-z)^{2})
\\
\nonumber &+&\int_{R_{0}^{2}}^{1/Q^{2}}d^{2}r m_{q}^{2}\int_{0}^{\dfrac{1}{r^{2}Q^{2}}}dz
\\
\nonumber &+& \int_{1/ \mu^{2}}^{R_{0}^{2}}d^{2}r\int_{0}^{\dfrac{1}{r^{2}Q^{2}}}dz 4Q^{2}z^{2}(1-z)^{2}(\dfrac{r^{2}}{R_{0}^{2}(x)})^{2}
\\
\nonumber &+& \int_{R_{0}^{2}}^{1/Q^{2}}d^{2}r\int_{0}^{\dfrac{1}{r^{2}Q^{2}}}dz 4Q^{2}z^{2}(1-z)^{2} \Big \},
\\
\nonumber &=&\dfrac{3\alpha_{em}}{32\pi^{2}}\sigma^{2}_{0}\sum_{q} e_{q}^{2}\big \{ \dfrac{-83}{90}+\dfrac{2}{R_{0}^{2}Q^{2}}
\\
\nonumber &+&\dfrac{1}{R_{0}^{4}Q^{4}}(\dfrac{1}{6}+\dfrac{1}{3}log(\mu^{2}R_{0}^{2})-\dfrac{4\mu^{2}}{3Q^{2}}+\dfrac{2\mu^{4}}{5Q^{4}})
\\
\nonumber &+&\dfrac{8}{9R_{0}^{6}Q^{6}}-\dfrac{1}{5R_{0}^{8}Q^{8}}+\dfrac{m_{q}^{2}}{2Q^{2}}(1+log(\dfrac{1}{Q^{2}R_{0}^{2}}) \big \},
\\
\end{eqnarray}
so the universal function by ignoring the logarithmic sentence is \\
\begin{eqnarray}
\nonumber g(\tau)&=&\dfrac{3\alpha_{em}}{32\pi^{2}} \sum_{q} e_{q}^{2}\big \{ \dfrac{-83}{90}+\dfrac{2}{\tau}+\dfrac{1}{\tau^{2}}(\dfrac{1}{6}-\dfrac{4\mu^{2}}{3Q^{2}}
\\
\nonumber &+& \dfrac{2\mu^{4}}{5Q^{4}})+\dfrac{8}{9\tau^{3}}-\dfrac{1}{5\tau^{4}}+\dfrac{m_{q}^{2}}{2Q^{2}} \big \}.
\\
\end{eqnarray}
The ratio $\dfrac{g(\tau)}{\alpha_{em}}$ for the universal functions (23) and (25) in terms of the scaling variable has been plotted
 in Fig. 2 in the different $x$ values for light quarks. According to these diagrams we can result the diffractive contribution
 of asymmetric pairs is dominated in the saturation limit so for $ 0< \tau <1 $ the system connected to a heavily absorbed diffractive event. \\

\section {\textbf{3. The contribution of heavy quarks in the diffractive process}}

In high energy the heavy flavors are useful in the agreement with the experimental data. A common way for study
 on the heavy productions has been based on the ratio method which is associated with the geometrical scaling [21].
 In pervious section we obtained the diffractive cross section quantitatively and showed by plotting Figs. 1 and 2 the
  geometrical scaling is established for light quarks. Now we assume the summation over flavors expands to include the charm
  quark with $m_{c}=1.5~ \mathrm{GeV}$ and plot the corresponding fraction $\dfrac{g(\tau)}{\alpha_{em}}$ in Fig. 3. According to these diagrams we see
there is not the dependence on $x$ and all curves almost fall on one, in the other words the geometrical scaling is confirmed.
The main cause of the increase in the value of this function is that the charm quark  because of its high mass is saturated at a higher order. \\
Fig. 4 by considering the bottom quark with
$m_{b}=4.75~\mathrm{GeV}$ to active flavors has been plotted that
shows these diagrams behave similar
to ones in Figs. 1 and 3 so the geometrical scaling is fulfilled.\\
 We note, since the heavy production appears in the feature of a symmetric dipole, universal functions (16) and (18)
 in plotting diagrams are used.\\
According to H1 and ZEUS reports the charm component of the structure function includes a significant fraction of the
proton structure function [22,23]. We calculate the contribution of this flavor in the diffractive cross section
\begin{equation}
\dfrac{\dfrac{d\sigma_{c}^{D}}{dt}\vert_{t=0} }{\dfrac{d\sigma ^{D}}{dt}\vert_{t=0}}=\dfrac{ g_{c}(\tau) }{g(\tau)}=\dfrac{\delta_{qc}\bigg ( g(\tau) \bigg )}{g(\tau)},
\end{equation}
the $\delta_{qc}$ function chooses charm quark from all active
flavors. $c\overline{c}$ pair is dominated in small size dipoles
with $r<R_{0}$ so
\begin{eqnarray}
\nonumber & & \dfrac{\dfrac{d\sigma_{c}^{D}}{dt}\vert_{t=0} }{\dfrac{d\sigma ^{D}}{dt}\vert_{t=0}}=\dfrac{ g_{c}(\tau) }{g(\tau)}=
\\
\nonumber & &\dfrac{e_{c}^{2}(\dfrac{17}{45}+\dfrac{m_{c}^{2}}{3Q^{2}})}{e_{c}^{2}(\dfrac{17}{45}+\dfrac{m_{c}^{2}}{3Q^{2}})+(e_{u}^{2}+e_{d}^{2}+e_{s}^{2})(\dfrac{17}{45}+\dfrac{(0.140)^{2}}{3Q^{2}})}.
\\
\end{eqnarray}
\\
The average value of the charm production in the diffractive
process by assuming $ 0.037\leq r_{c}\leq 0.13~  \mathrm{fm} $ is
about 40 percent according
to Fig. 5. This figure shows the corresponding fraction is independent of $x$ and $Q^{2}$ and all curves fall on one. \\
We can calculate this fraction for estimating of the bottom production $b\bar{b}$ as
\begin{equation}
\dfrac{\dfrac{d\sigma_{b}^{D}}{dt}\vert_{t=0} }{\dfrac{d\sigma ^{D}}{dt}\vert_{t=0}}=\dfrac{ g_{b}(\tau) }{g(\tau)}=\dfrac{\delta_{qb}\bigg ( g(\tau) \bigg )}{g(\tau)},
\end{equation}
$\delta_{qb}$ selects the bottom quark from five active flavors.
Therefore the possibility for finding the bottom production becomes
  \begin{eqnarray}
\nonumber & & \dfrac{\dfrac{d\sigma_{b}^{D}}{dt}\vert_{t=0} }{\dfrac{d\sigma ^{D}}{dt}\vert_{t=0}}=\dfrac{ g_{b}(\tau) }{g(\tau)}=
\\
\nonumber & &\dfrac{e_{b}^{2}(\dfrac{17}{45}+\dfrac{m_{b}^{2}}{3Q^{2}})}{e_{b}^{2}(\dfrac{17}{45}+\dfrac{m_{b}^{2}}{3Q^{2}})+e_{c}^{2}(\dfrac{17}{45}+\dfrac{m_{c}^{2}}{3Q^{2}})+\dfrac{2}{3}(\dfrac{17}{45}+\dfrac{(0.140)^{2}}{3Q^{2}})},
\\
\end{eqnarray}
\\
that is about $10$ percent in the range $ 0.014\leq r_{b}\leq
0.043~  \mathrm{fm} $ that is seen from Fig.6. In recent relations
the important element is $(\dfrac{m_{q}^{2}}{Q^{2}})$ that along
with $ e_{q}^{2}$ includes inherent characterizes of the dipole.
Also all of curves in Fig. 6 fall on one that means there is not dependence on $x$ and $Q^{2}$.\\
To continue we can take a step forward and obtain the ratio of the diffractive cross section to the total cross section [20].
For symmetric dipoles with $r < R_{0}$ we get\\
  \begin{eqnarray}
\nonumber & &R(\tau)=\dfrac{\dfrac{d\sigma^{D}_{tot}}{dt}\vert_{t=0}}{\sigma^{\gamma^{\star}P}_{tot} (x,Q^{2})}=\dfrac{\sigma_{0}^{2}g(\tau)}{\sigma_{0}f(\tau)}
\\
\nonumber & & = \dfrac{\sigma_{0}}{16\pi \tau} \Bigg ( \dfrac{\sum_{q} e_{q}^{2}(\dfrac{17}{45}+\dfrac{m_{q}^{2}}{3Q^{2}})}{\sum_{q} e_{q}^{2}(\dfrac{11}{15}+\dfrac{m_{q}^{2}}{2Q^{2}})} \Bigg ),
\\
\end{eqnarray}
\\
and where $r > R_{0}$ we will have
 \begin{eqnarray}
\nonumber & &R(\tau)=\dfrac{\dfrac{d\sigma^{D}_{tot}}{dt}\vert_{t=0}}{\sigma^{\gamma^{\star}P}_{tot} (x,Q^{2})}=\dfrac{\sigma_{0}^{2}g(\tau)}{\sigma_{0}f(\tau)}
\\
\nonumber & & = \dfrac{\sigma_{0}}{16\pi } \Bigg ( \dfrac{\sum_{q} e_{q}^{2}(\dfrac{7}{15}-\dfrac{4\tau}{45}+\dfrac{m_{q}^{2}}{Q^{2}}(1-\dfrac{2\tau}{3}))}{\sum_{q} e_{q}^{2}(\dfrac{4}{5}-\dfrac{4\tau}{60}+\dfrac{m_{q}^{2}}{Q^{2}}(1-\dfrac{\tau}{2}))} \Bigg ).
\\
\end{eqnarray}
\\
These ratios only relate to the size, mass and charge of color dipoles. We have plotted them in Fig. 7
in terms of the scaling variable, $\tau$, for light flavors. In $ 0 < \tau < 1 $ we see a flat function
 that for $ \tau > 1 $ reduces proportionally to $\dfrac{1}{\tau}$  that express in the diffractive event
 the saturation is dominated and the interaction in the scaling region occurs rarely. It is necessary to mention there is
 no dependence on $x$ and $Q^{2}$ [24-26]. We expect this behavior also remains by adding the contribution of heavy quarks.\\
 Our other suggestion for showing the ratio of the diffractive cross section to the total cross section is independent of $x$
 and $Q^{2}$ so we plot $\dfrac{R(\tau)}{\sigma_{0}}$ in terms of  the center of mass energy of $\gamma^{\star}p$ , $W$, as [12].
 This quantity in Fig. 8 has been plotted for light quarks with $m_{q}=140 MeV$ for different $x$ values.
 We see a flat function for $W>100~ \mathrm{GeV}$ that indicates the invariance of this ratio.\\

\section {\textbf{4. summary}}

The color dipole picture is an effective field theory in describing the small-$x$ limit of QCD without nonlinear sentences in the evolution equation
and connects to the unitarity effect due to the gluon recombination. In this analysis we discuss the diffractive deep inelastic scattering by
 the color dipole picture. We believe our model represents the basic dynamics because it allows us to study a wide range of data in a satisfactory way.
 We can assume by following Good and Walker method diffractive eigenstates as colorless quark-antiquark pairs which remain unchanged during the scattering.The
 diffractive process is characterized by a final state in which a large rapidity gap is not filled with particles. The LRG in the limit of the unitarity
 saturation may be terminated by the absorptive correction.\\
We came to the conclusion that in both the symmetric and
asymmetric dipoles the diffraction is sensitive to the saturation
effect since the diffractive cross section is proportional to $
\sigma^{2}_{0} $. Fig. 1 shows there is a smooth transfer from the
color transparency to the saturation when the scaling variable is
around $1$. Also in the enough energy by adding the contribution
of heavy flavore this transition in Figs. 3 and 4 is seen and the
small-$x$ saturation is proven.
According to Fig. 2 we conclude that extrapolation to the diffraction process proves the saturation effect well for large-size dipoles
and shows the interaction in the scaling region occurs rarely.\\
The probability of the charm production in $x<0.01$ directly originated from virtual photon in the
diffractive process is about 40 percent. This impressive contribution expresses the importance of the charm cross section in colliders in the high energy.
The corresponding fraction for the bottom production decreases to 10 percent because of small size and large mass of this flavor.
From Figs. 5 and 6 we calculate the geometrical scaling is confirmed in the diffractive process including heavy flavores and both of the obtained
magnitudes depends on the mass, size and charge of the involved active quarks.\\
The ratio of the diffractive cross section to the total cross section painted in Fig. 7 is the other quantity depends on inherent
 characteristics of dipoles and remains unchanged relative to $x$ variable. Fig. 8  is another confirmation to show this ratio is invariance.\\
The significant conclusion of this work is that the diffractive event in the color dipole model probes QCD in a different way, for instance, the
unitarity is an important ingredient associated with the saturation effect that leads to a good description of data.\\
 In conclusion, the idea of the geometrical scaling for the diffractive cross section is in place. The universal functions
 obtained are not bounded functions and only depends on the inherent properties of dipoles.\\

\newpage
\section{References}
1.C. Ewerz and O. Nachtmann, Annals Phys. $\mathbf{322}$, 1635 (2007) [arXiv:hep-ph/0404254].\\
2.C. Ewerz and O. Nachtmann, Annals Phys. $\mathbf{322}$, 1670 (2007) [arXiv:hep-ph/0604087].\\
3.H. Kowalski and A. Luszczak, Phys. Rev. $\mathbf{D89}$, 074051 (2014).\\
4.A. Luszczak and H. Kowalski, Phys. Rev. $\mathbf{D953}$, 014030 (2017).\\
5.K. Golec-Bierant, Acta Phys. Polon. $\mathbf{B53}$ (2004)[arXiv:hep-ph/0311278v2].\\
6.C. Marquet and L. Schoeffel, Phys. Lett. $\mathbf{B639}$ (2008) [arXiv:hep-ph/0606079].\\
7.Stanley J. Brodsky, Ivan Schmidt and Jian Yang, Phys. Rev. $\mathbf{D10}$, 116003 (2004) [arXiv:hep-ph/0409279].\\
8.R.J. Glauber, Phys. Rev, $\mathbf{100}$, 242 (1955).\\
9.M.L. Good and W.D. Walker, Phys. Rev. $\mathbf{120}$, 1857 (1960).\\
10.V.P. Goncalves and M.V.T. Machado, Phys.Rev.Lett. $\mathbf{91}$, 202002 (2003) [arXiv:hep-ph/0307090].\\
11.B. Z. Kopeliovich, L. I. Lapidus and A. B. Zamolodchikov, JETP Lett. $\mathbf{33}$, 595 (1981).\\
12.J. Bartels, K. Golec-Bierant and H. Kowalski, Phys. Rev. $\mathbf{D66}$ (2002).\\
13.C. Marquet, Phys.Rev. $\mathbf{D 76}$, 094017 (2007).\\
14.H. Kowalski, T. Lappi, C. Marquet and R. Venugopalan,  Phys. Rev. $\mathbf{C78}$, 045201 (2008).\\
15.K. Golec-Biernat and M. W$\mathrm{\ddot{u}}$sthoff, Phys. Rev.
$\mathbf{D60}$, 114023 (1999)
[arXiv:hep-ph/9903358].\\
16.M.S. Kugeratski, V.P. Goncalves and F.S. Navarra, Eur. Phys. J. $\mathbf{C46}$ (2006) [arXiv:hep-ph/0511224].\\
17. S. Aid, et. al. [H1 Collaboration], Nucl. Phys. $\mathbf{B470}$, (1996); M.
Derrick, et. al. [ZEUS Collaboration], Z. Phys. $\mathbf{C72}$, 399 (1996).\\
18.K. Golec-Biernat and M. W$\mathrm{\ddot{u}}$sthoff, Phys. Rev. $\mathbf{D 59}$, 014017 (1999).\\
19.K. Golec-Biernat and M. W$\mathrm{\ddot{u}}$sthoff, Eur. Phys. J. $\mathbf{C 20}$, 313 (2001).\\
20.Z. Jalilian and G.R. Boroun, Phys. Lett. $\mathbf{B773}$, p.455-461 (2017).\\
21.T. Steble, Phys. Rev. $\mathbf{D88}$, 014026 (2013).\\
22.V.P. Goncalves, B. Kopeliovich, J. Nemchik, R. Pasechnik and I. Potashnikova, Phys. Rev. $\mathbf{D96}$, 014010 (2017).\\
23.K. Werner, B. Goiot, lu. Karpenko, T. Pierog and G. Sophys, Conference: C15-11-23, p.66-70, (2016) [arXiv:hep-ph/1602.03414v1]\\
24.K. Golec-Biernat and M. W$\mathrm{\ddot{u}}$sthoff, Phys. Rev. $\mathbf{D59}$, 014017 (1998).\\
25.C. Adloff et. al. [H1 Collaboration],  Z. Phys. $\mathbf{C76}$, 6130 (1997).\\
26.J. Braitwey, et. al. [ZEUS Collaboration], Eur. Phys. J. $\mathbf{C1}$, 810 (1998).\\



\begin{figure}
\includegraphics[width=0.7\textwidth]{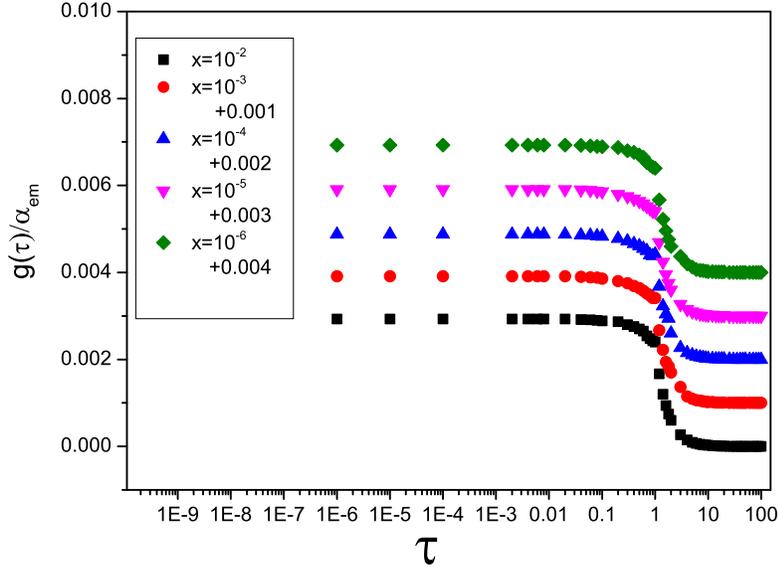}
\caption{The ratio $\dfrac{g(\tau)}{\alpha_{em}}$ for symmetric pairs in the different $x$
values belong to the range $10^{-6}-10^{-2}$ to $\tau$ variable for light flavors. }\label{Fig1}
\end{figure}

\begin{figure}
\includegraphics[width=0.7\textwidth]{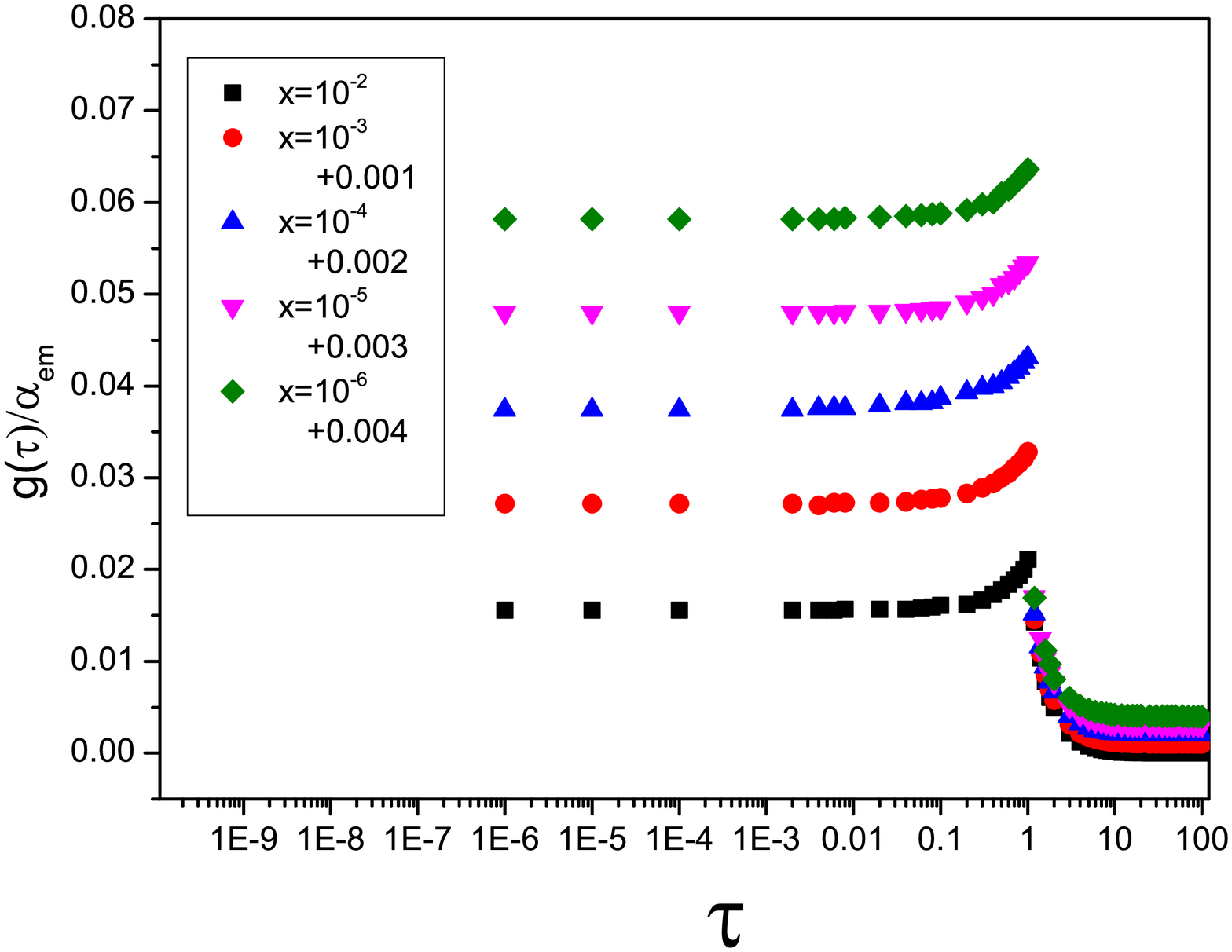}
\caption{The ratio $\dfrac{g(\tau)}{\alpha_{em}}$ for asymmetric pairs in the different
$x$ values belong to the range $10^{-6}-10^{-2}$ to $\tau$ variable for light flavors. }\label{Fig2}
\end{figure}

\begin{figure}
\includegraphics[width=0.7\textwidth]{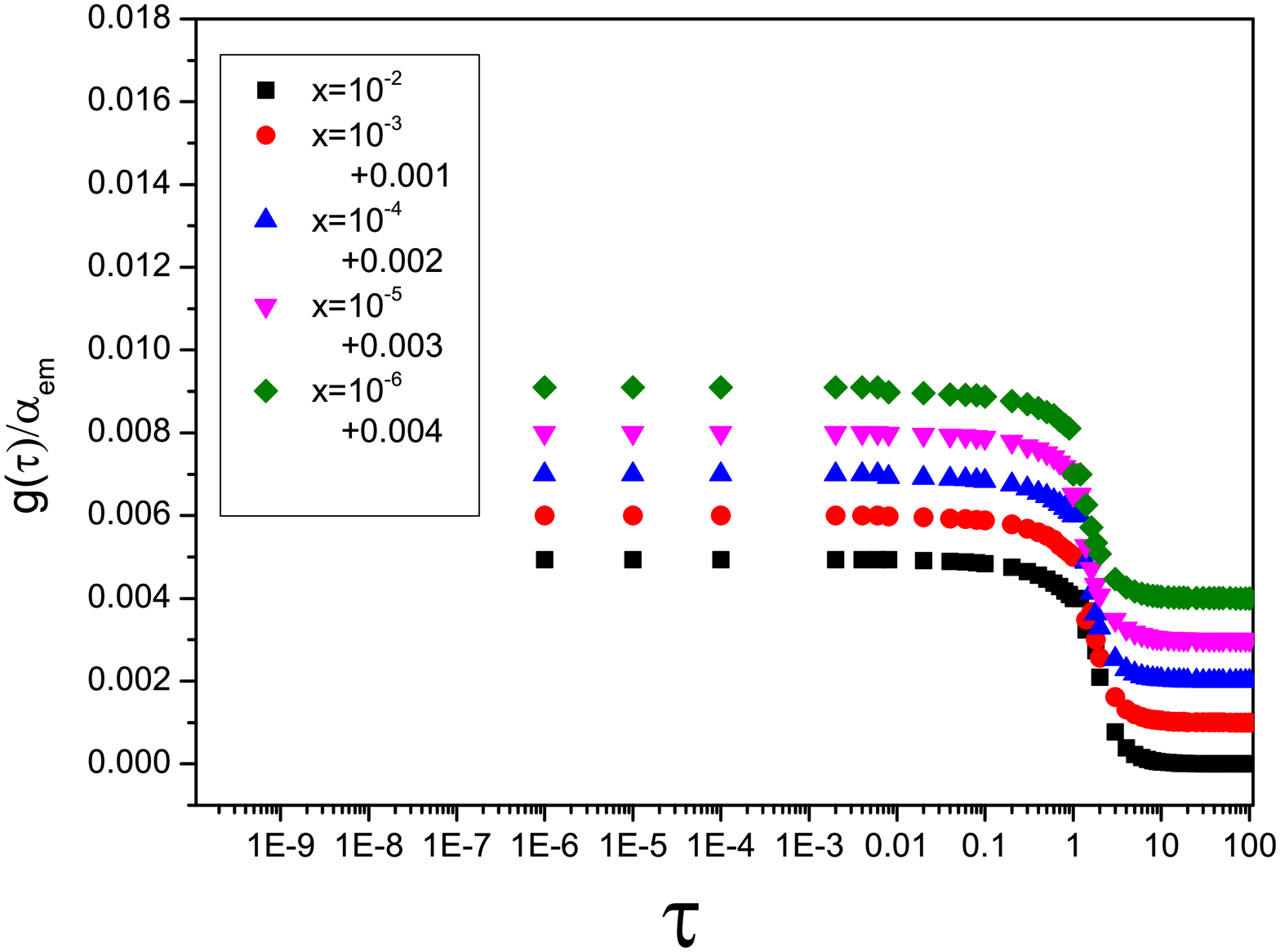}
\caption{The ratio $\dfrac{g(\tau)}{\alpha_{em}}$ by considering light and charm flavors
of symmetric dipoles in the different $x$ values belong to the range $10^{-6}-10^{-2}$ to $\tau$ variable. }\label{Fig3}
\end{figure}

\begin{figure}
\includegraphics[width=0.7\textwidth]{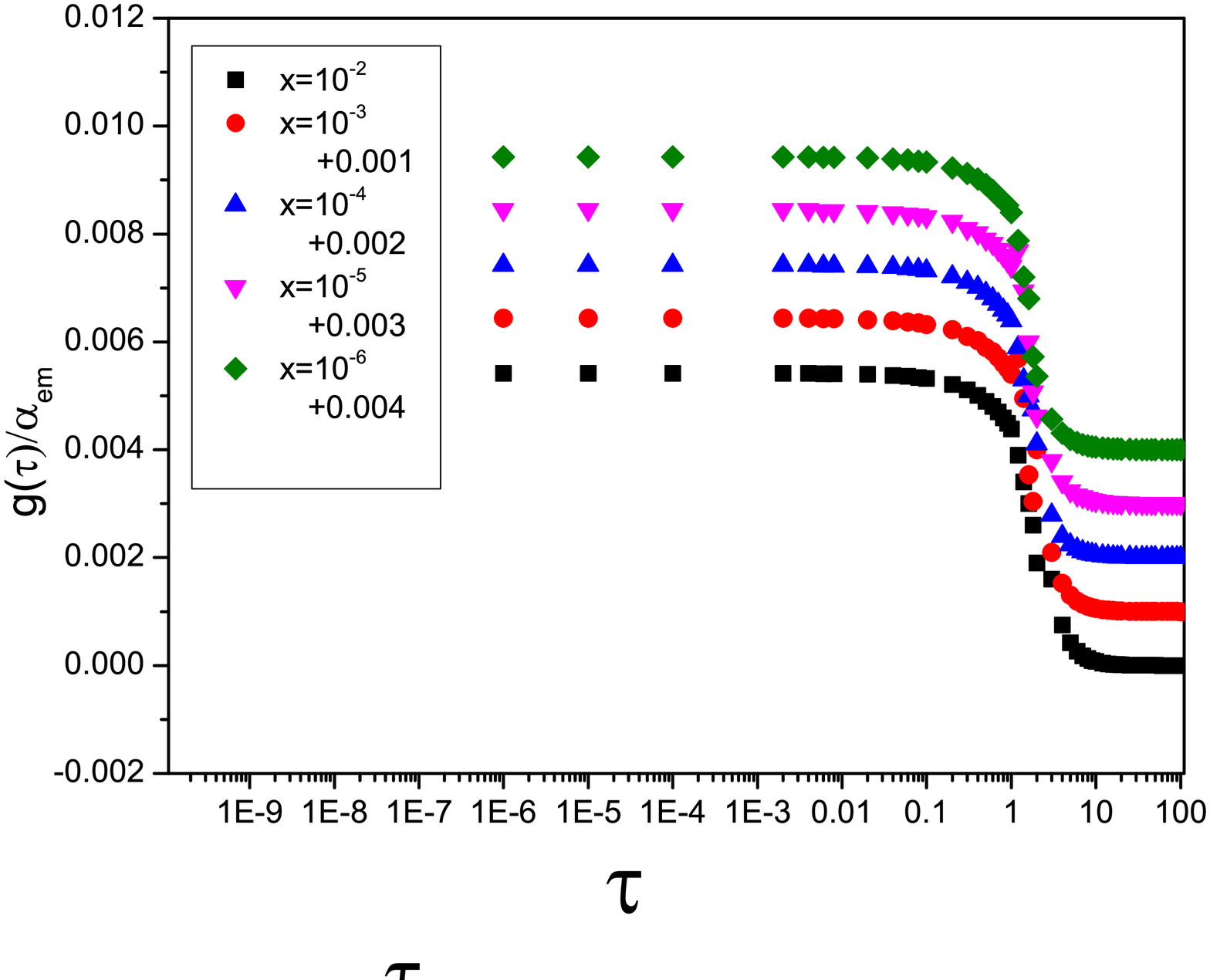}
\caption{The ratio $\dfrac{g(\tau)}{\alpha_{em}}$ by considering
light, charm and bottom flavors of symmetric dipoles in the
different $x$ values belong to the range $10^{-6}-10^{-2}$ to
$\tau$ variable. }\label{Fig4}
\end{figure}

\begin{figure}
\includegraphics[width=0.7\textwidth]{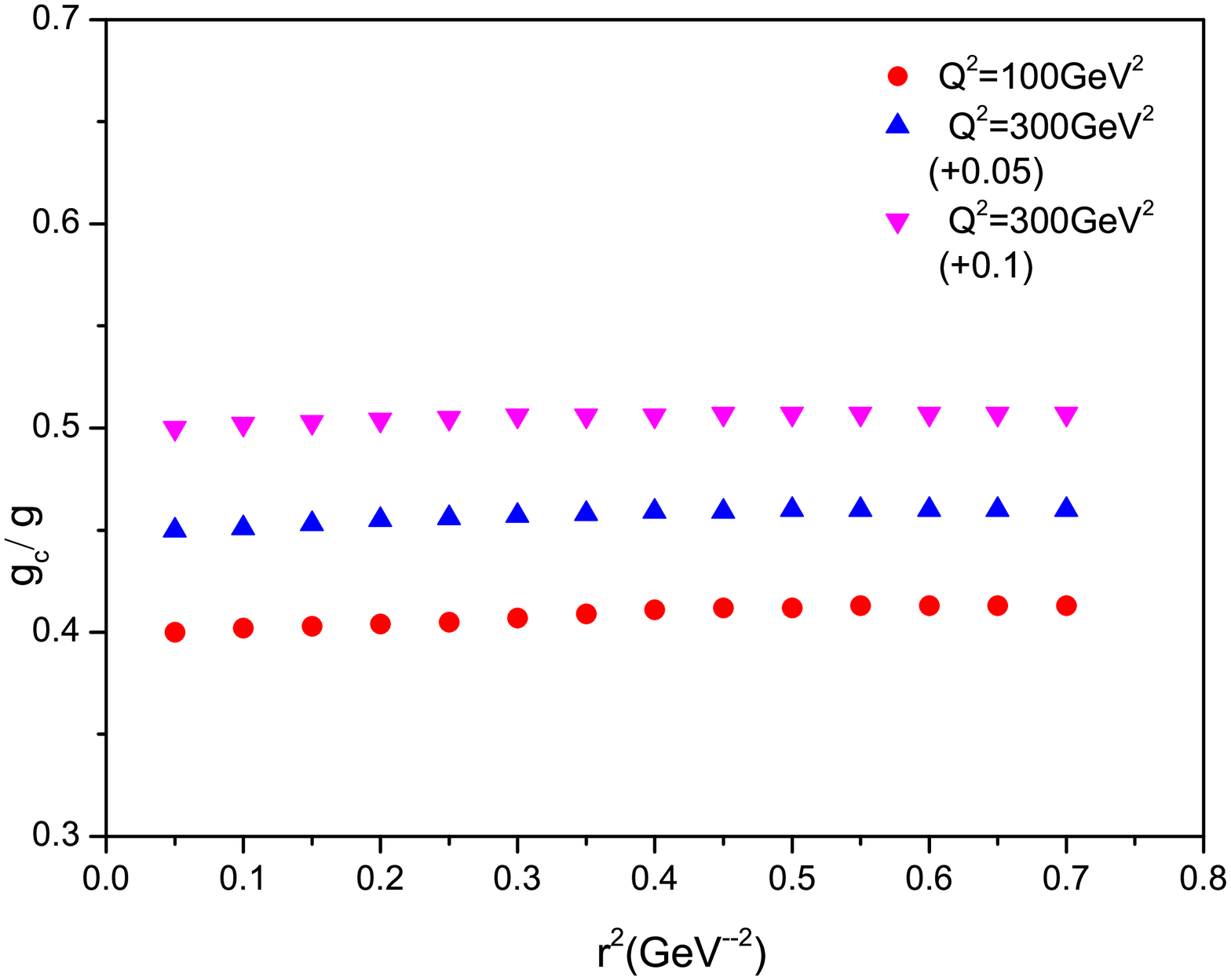}
\caption{The contribution of the charm quark in the diffractive
cross section for $Q^{2}=100, ~200 ~\mathrm{and} ~300~
\mathrm{GeV}^{2}$.}\label{Fig6}
\end{figure}

\begin{figure}
\includegraphics[width=0.7\textwidth]{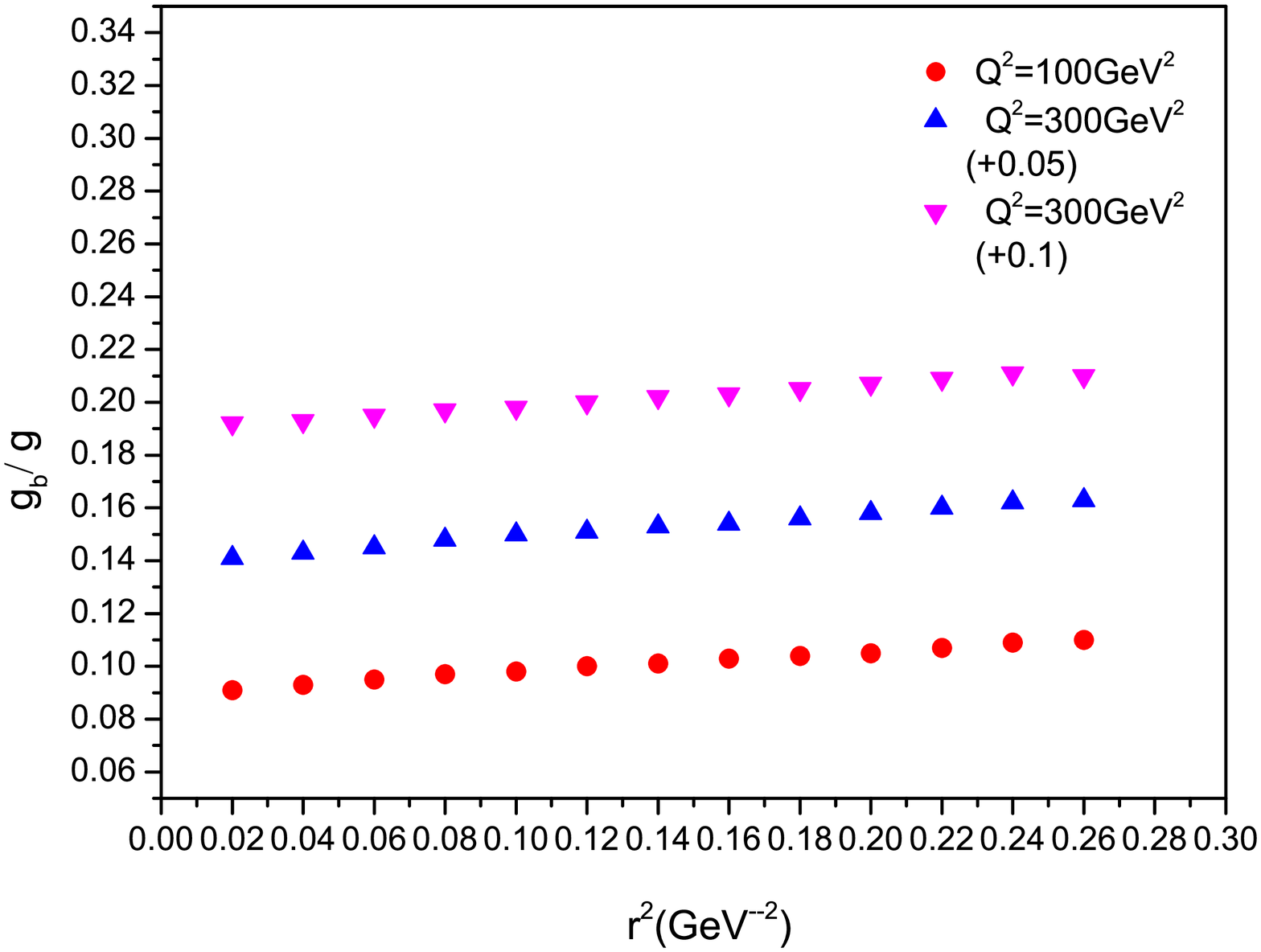}
\caption{The contribution of the bottom quark in the diffractive
cross section for $Q^{2}=100, ~200 ~\mathrm{and} ~\mathrm{300}~
\mathrm{GeV}^{2}$.}\label{Fig6}
\end{figure}

\begin{figure}
\includegraphics[width=0.7\textwidth]{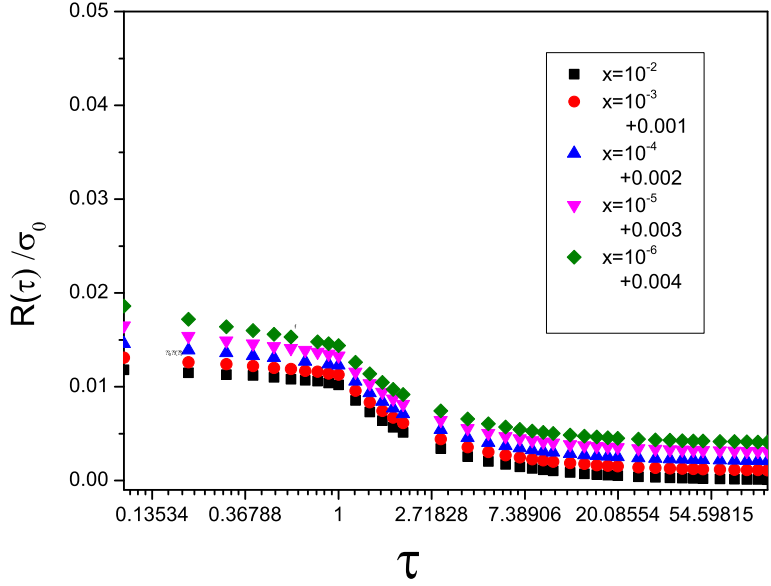}
\caption{The ratio $\dfrac{1}{\sigma_{0}} \dfrac{\dfrac{d\sigma^{D}_{tot}}{dt}\vert_{t=0}}{\sigma^{\gamma^{\star}P}_{tot}}=\dfrac{R(\tau)}{\sigma_{0}}$
in the different $x$ values belong to the range $10^{-6}-10^{-2}$ to $\tau$ variable for light flavors.}\label{Fig7}
\end{figure}

\begin{figure}
\includegraphics[width=0.7\textwidth]{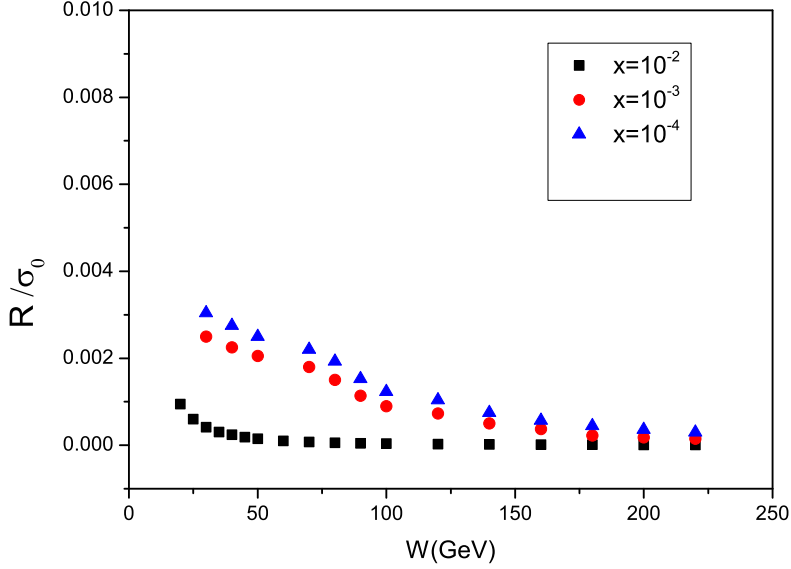}
\caption{The ratio $\dfrac{1}{\sigma_{0}} \dfrac{\dfrac{d\sigma^{D}_{tot}}{dt}\vert_{t=0}}{\sigma^{\gamma^{\star}P}_{tot}}=\dfrac{R}{\sigma_{0}}$
in the different $x$ values  to the center of mass energy of $\gamma^{*}p$ for light flavors.}\label{Fig8}
\end{figure}


\end{document}